\documentclass{iopart}
\begin{document}
\hspace*{4 in}CUQM-122\\


\title[Eigenvalue bounds for polynomial central potentials in $d$ dimensions]{Eigenvalue bounds for polynomial central potentials in $d$ dimensions}
\author{Qutaibeh D. Katatbeh}
\address{
Department of Mathematics and Statistics,
Faculty of Science and Arts,
Jordan University of Science and Technology, 
Irbid, Jordan 22110
}\ead{qutaibeh@yahoo.com}
\author{Richard L. Hall}%
\address{Department of Mathematics and Statistics, Concordia University,
1455 de Maisonneuve Boulevard West, Montr\'eal,
Qu\'ebec, Canada H3G 1M8} \ead{rhall@mathstat.concordia.ca}
\author{Nasser Saad}
\address{Department of Mathematics and Statistics,
University of Prince Edward Island,
550 University Avenue, Charlottetown,
PEI, Canada C1A 4P3.}\ead{nsaad@upei.ca}
\def\dbox#1{\hbox{\vrule  
                        \vbox{\hrule \vskip #1
                             \hbox{\hskip #1
                                 \vbox{\hsize=#1}%
                              \hskip #1}%
                         \vskip #1 \hrule}%
                      \vrule}}
\def\qed{\hfill \dbox{0.05true in}}  
\def\square{\dbox{0.02true in}} 
\begin{abstract} If a single particle obeys non-relativistic QM in ${\bf R}^d$ and has the Hamiltonian
$H=- \Delta+ f(r),$ where $f(r)=\sum_{i = 1}^{k}a_ir^{q_i},~ 2\leq q_i < q_{i+1},~ a_i \geq 0$, then the eigenvalues $E=E_{n\ell}^{(d)}(\lambda)$ are given approximately by the semi-classical expression
$E = \min\limits_{r > 0}\left\{\frac{1}{r^2} + \sum_{i = 1}^{k}a_i(P_ir)^{q_i}\right\}.$ It is proved that this formula yields a lower bound if $P_i  = P_{n\ell}^{(d)}(q_1)$, an upper bound if $P_i = P_{n\ell}^{(d)}(q_k)$ and a general approximation formula if $P_i = P_{n\ell}^{(d)}(q_i)$.  For the quantum anharmonic oscillator $f(r)=r^2+\lambda r^{2m},m=2,3,\dots$ in $d$ dimension, for example, $E=E_{n\ell}^{(d)}(\lambda)$ is determined by the algebraic expression 
$\lambda={1\over \beta}\left({2\alpha(m-1)\over mE-\delta}\right)^m\left({4\alpha \over (mE-\delta)}-{E\over (m-1)}\right)
$
where
$\delta={\sqrt{E^2m^2-4\alpha(m^2-1)}}$ and $\alpha , \beta$  are constants. An improved lower bound to the lowest eigenvalue in each angular-momentum subspace is
also provided. A comparison with the recent results of Bhattacharya et al (Phys. Lett. A, 244 (1998) 9) and Dasgupta et al (J. Phys. A: Math. Theor., 40 (2007) 773) is discussed.
\end{abstract}
\vspace{2pc}
\maketitle
\section{Introduction and main results}
\noindent The purpose of the present work is to establish a global bound formula for the discrete spectrum $\{E_{n\ell}^{(d)}\}$, $n=1,2,\dots, l=0,1,2,\dots$ of the $d$-dimension Schr\"odinger equation with polynomial potentials given by
\begin{equation}\label{eq1}
H\psi =\left(- \Delta+ \sum_{i = 1}^{k}a_ir^{q_i}\right)\psi=E\psi,~~2\leq q_i < q_{i+1},
\end{equation}
where $\Delta$ is the $d$-dimensional Laplacian operator, $r=\|{\mathbf r}\|, {\mathbf r}\in {\bf R}^d$, and the coefficients $a_i \geq 0,$ are not all zero. The key motivation for our present study lies in the well-known fact that the majority of quantitative predictions of Schr\"odinger's equation with a polynomial potential (\ref{eq1}) in nuclear, atomic, molecular, and condensed matter physics must usually rely on numerical estimates \cite{znojil}-\cite{campy}. Thus, a simple global eigenvalue formula can serve as a basis for exploration and also for checking different approximate methods in quantum mechanics \cite{ran}. Another important motivation for the present work is a recent contribution by Dasgupta {\it et al} \cite{dasgupta} regarding a general simple scheme for evaluating the ground state as well the excited-state energies for $\lambda r^{2m}$ quantum anharmonic oscillators in one dimension, see also \cite{ran}. We provide in the present work a more general scheme sufficient to generate all energy levels in arbitrary dimension, not only of $r^{2m}$ quantum anharmonic oscillators, but also for any polynomial potential of the form $\sum_{i = 1}^{k}a_ir^{q_i}$ a sufficient degree of accuracy to be interesting. The purpose is not merely to obtain accurate energy eigenvalues for different polynomial potentials for which a large number of methods exist in the literature. Rather, we propose a simple approach which provides energy bounds as well as an approximate energy formula with a reasonable accuracy and with a minimum amount of effort. Consider, as an example, the celebrated quantum anharmonic oscillator \cite{simon}-\cite{weniger} Hamiltonian $-\Delta +r^2+\lambda r^{2m}, m=2,3,\dots$ in $d$ dimensions: we show that for any state $n=1,2,\dots$, the eigenenergy $E=E_{n\ell}^{(d)}(\lambda)$ is determined approximately by the expression
\begin{equation}\label{eq2}
\lambda={1\over \beta}\left({2\alpha(m-1)\over mE-\delta}\right)^m\left({4\alpha \over (mE-\delta)}-{E\over (m-1)}\right)
\end{equation}
where
$\delta={\sqrt{E^2m^2-4\alpha(m^2-1)}}$ and $\alpha$ and $\beta$ are constants. Further, we show that upper or lower  bounds for the energy eigenvalues (\ref{eq2}) for a given state are expressed in terms of a single constant for any value of $\lambda$.  The dependence of $\alpha$ and $\beta$ on $m$ and $d$ will be discussed in a subsequent section.
We obtain our global eigenvalue formula for (\ref{eq1}) by using the so-called $P$-representation \cite{hall89} for the Schr\"odinger spectra generated by the pure power-law potential ($q>0$). In this representation, a discrete eigenvalue $\epsilon$ is written as the minimum of a function of one variable $r$ and a parameter $P$: this induces a one-one relation between $\epsilon$ and $P.$ More specifically, we write
\begin{equation}\label{eq3}
(-\Delta + r^q)\psi_{n\ell}^{(d)}= \epsilon_{n\ell}^{(d)}(q)\psi_{n\ell}^{(d)}\Rightarrow \epsilon_{n\ell}^{(d)}(q) = \min_{r>0}\left[\frac{1}{r^2} + \left(P_{n\ell}^{(d)}(q)r\right)^q\right],
\end{equation}
where
\begin{equation}\label{eq4}
P_{n\ell}^{(d)}(q) = \left[\epsilon_{n\ell}^{(d)}(q)\right]^{(2+q)/2q} \bigg[{2\over 2+q}\bigg]^{1/q}  \bigg[{q\over 2+q}\bigg]^{1/2}.
\end{equation}
This may seem at first sight rather inconvenient since the computation of $P$ requires knowledge of $\epsilon$. An important advantage of (\ref{eq4}), however, is that the computation of $P$ is independent of the potential parameters. In other words, the computation of $P$ for $H=-\Delta + r^q$ is sufficient to yield the discrete spectrum of the Hamiltonian $H_v=-\Delta + v r^q$ with eigenvalues given by
\begin{equation}\label{eq5}
 E_{n\ell}^{(d)}(q) = \min_{r>0}\left[\frac{1}{r^2} + v \left(P_{n\ell}^{(d)}(q)r\right)^q\right]
\end{equation}
for arbitrary $v>0$. This may seem unnecessary for a Hamiltonian of the form $H_v$ because a simple scaling argument shows $E_{n\ell}^{(d)}(q;v)=v^{2/(q+2)} E_{n\ell}^{(d)}(q)$; but for polynomial potentials, as in (\ref{eq1}), where $a_i>0, i=1,2,\dots, k$, equations (\ref{eq3})-(\ref{eq4}) play an important role in establishing some of the general energy formulae \cite{hall92} through the decomposition of the Hamiltonian (\ref{eq1}) by means of 
\begin{equation}\label{eq6}
H=- \Delta+ \sum_{i = 1}^{k}a_ir^{q_i}= \sum_{i = 1}^{k}\omega_i H^{(i)}
\end{equation}
where
\begin{equation}\label{eq7}
H^{(i)}=-\Delta+{a_i\over \omega_i}r^{q_i}.
\end{equation}
and $\{\omega_i\}_{i=1}^k$ is an arbitrary set of positive weights with sum equal to 1.
Further, it worth mentioning that this dependence can be resolved for certain special values of $q$, for example if $q=2$, we know that \cite{hall2002}
\begin{equation}\label{eq8}
P_{n\ell}^d(2)\Rightarrow\left\{ \begin{array}{rl}
 P_{n\ell}^d(2)=(2n+l+{d\over 2}-2) &\mbox{ if $d\geq 2$} \\
\\
P_n(2)=(n-{1\over 2})\quad\quad\quad\quad&\mbox{ if $d=1$}
       \end{array} \right.
\end{equation}
The main results of the present work may be summarized by the following two theorems:
\medskip

\noindent{\bf Theorem A:} \emph{Eigenvalue bounds for the spectrum $\{E_{n\ell}^{(d)}\}$ of the Hamiltonian (\ref{eq1}) are given by
\begin{equation}\label{eq9}
{\mathcal E} \equiv \min_{r > 0}\left[\frac{1}{r^2} + \sum_{i = 1}^{k}a_i(P_ir)^{q_i}\right].
\end{equation}
where
\begin{enumerate}
  \item[i)]  if $P_i  = P_{n\ell}^{(d)}(q_1)$, then $\mathcal{E} \leq E_{n\ell}^{(d)}$
  \item[ii)] if $P_i = P_{n\ell}^{(d)}(q_k)$ then $\mathcal{E} \geq E_{n\ell}^{(d)}$
    \end{enumerate}
Here the numbers $P_{n\ell}^{(d)}(q_i)$ are given by (\ref{eq4}).
}
\vskip 0.1true in
\noindent{\bf Theorem B:} 
\emph{The eigenvalues of the Hamiltonian (\ref{eq1}) are given approximately by the semiclassical formula
\begin{equation}\label{eq10}
 {\mathcal E} \equiv \min_{r > 0}\left[\frac{1}{r^2} + \sum_{i = 1}^{k}a_i(P_ir)^{q_i}\right].
\end{equation} 
where, for the lowest eigenvalue in each angular-momentum subspace, and $d\geq 1$, we have
\begin{enumerate}
\item[i)] For $n=1$ and $P_i = P_{1\ell}^{(d)}(q_i), i=1\dots k$, then $\mathcal{E} \leq E_{1\ell}^{(d)}$ for all $d\geq 1$ and $l=0,1,2,\dots$.
\item[ii)] For $n\geq 2$ and $P_i = P_{n\ell}^{(d)}(q_i), i=1\dots k$, $\mathcal{E} \approx E_{n\ell}^{(d)}$. 
\end{enumerate}
Further, for the lowest eigenvalue in $d\geq 1$, we have
\begin{enumerate}
\item[iii)] ${\mathcal E}\leq E_{10}^{(d)}$ if the numbers $P_i$   are replaced by the  explicit lower approximations for $P_{10}^{(d)}(q_i)$ given by
\begin{equation}\label{eq11}
P_i = \left({de\over 2}\right)^{1\over 2}\left({d\over {q_i e}}\right)^{1\over q_i}\left[{\Gamma[1+{d\over 2})\over \Gamma(1+{d\over q_i})}\right]^{1\over d},\quad e=\exp(1)
\end{equation}
\item[iv)] ${\mathcal E}\geq E_{10}^{(d)}$ if the numbers $P_i$ are replaced by the explicit upper approximations to $P_{10}^{(d)}(q_i)$ given by
\begin{equation}\label{eq12}
P_i =\left({d\over 2}\right)^{1\over 2}\left[ {\Gamma({d+q_i\over 2})\over \Gamma({d\over 2})}\right]^{1\over q_i}.  \end{equation}
\end{enumerate}
}
\noindent The difference between the two parts of Theorem B is that, in the first part (i)-(ii), the $P$-numbers are to be computed from the pure-power energies by use of (\ref{eq4}), whereas, in the second part (iii)-(iv), the $P$-numbers are given explicitly in terms of the Gamma function.  We use the term ``semiclassical'' in the following sense: once the component kinetic potentials have been fixed by the $P$-numbers, what remains is a minimization over a real function; in the approximation, this expresses the trade off between the kinetic and potential energies; the final picture is semiclassical since the kinetic energy is reduced to $1/r^2$ and a wave equation is no longer involved.
\vskip0.2true in
\noindent In the next section, we discuss the proof of these two theorems. In section (3), the application of these two theorems to  the quantum anharmonic oscillator Hamiltonian is presented. The conclusion is given in section~4.
\section{Proof of theorems A and B}
\subsection{Proof of Theorem A} 
\noindent The proof of theorem A depends on the application of envelope theory and Kinetic potentials technique developed earlier by Hall \cite{hall83}-\cite{hall93} and used successfully since then. We shall outline here a brief summary of the theory to provide us with sufficient details to prove the theorem, and we refer the interested reader to Ref. \cite{hall83}-\cite{hall93} for more details. For simplicity, we present this brief summary for the case of $d=3$ spatial dimensions: for arbitrary $d$, the extension is straightforward. Consider the  Schr\"odinger operators of the form 
\begin{equation}\label{eq13}
H=-\Delta+vf(r), 
\end{equation}
where $f$ is the shape of a central potential in ${\bf R}^3$, and $v>0$ is
the coupling parameter. The principal idea of envelope theory \cite{hall83}-\cite{hall84} is that the minimization of the Rayleigh quotient $(\psi,H\psi)/(\psi,\psi)$ is performed in two stages. The first stage, with $\langle\psi, -\Delta\psi\rangle =s$ fixed, involves only the shape  of the potential $f$ and leads to a family $\{\overline f_{n\ell}\}$ of kinetic potentials $\overline f_{n\ell}(s)$. Here $s$ is a positive constraint variable: it only becomes the mean kinetic energy when the minimization of the sum of the kinetic and potential energies has been effected.  We have
\begin{equation}\label{eq14}
E_{n\ell}=\min_{s>0}\{s+ v \overline{f}_{n\ell}(s)\} \end{equation}
in which the critical value of $s=\langle \psi,-\Delta\psi\rangle >0$ is the mean kinetic energy.  The kinetic potentials \cite{hall92}, which represent the result of min-max theory applied to the potential shape $f$ for fixed $s$, are given as a Legendre transformation $\{s = E(v) - vE'(v),\ \overline{f}(s) = E'(v)\}$ of the function $E(v)$, which describes how the eigenvalue depends on the coupling $v$. They may also be defined by the following general formula
\begin{equation}\label{eq15}
\overline{f}_{n\ell}(s)=\inf\limits_{{\cal D}_{n\ell}}\sup\limits_{{{\scriptstyle \psi \in {\cal D}_{n\ell}} \atop {\scriptstyle
\|\psi\| = 1}}}\int \psi(r)f([(\psi,-\Delta\psi)/s]^{1/2} r)\psi(r) d^3r,
\end{equation}
where ${\cal D}_{n\ell}$ is the span of a set of $n$ linearly independent functions. It is interesting to notice that the kinetic potential $\overline{f}_{n\ell}$ can be replaced by the potential $f(r)$ itself through the  parameterization of $\overline{f}_{n\ell}(s)$ in term of the variable $r$ (used here as a new parameter to replace $s$), that is to say $\overline{f}_{n\ell}(s)=f(r)$.  We now invert this monotone function to give the $K$ functions \cite{hall93}
\begin{equation}\label{eq16}
s=(\overline{f}_{n\ell}^{-1}\circ f)(r)=K_{n\ell}^{(f)}(r).
\end{equation}
It is easy to show that the $K$ functions obey the scaling property 
\begin{equation}\label{eq17}
Af\left({r\over b}\right) + B\rightarrow \left({1\over b^2}\right)K\left({r\over b}\right),
\end{equation}
and in general they are independent of coupling and potential shifts \cite{hall93}. The eigenvalues are recovered from the $K$ functions by the expression
\begin{equation}\label{eq18}
E_{n\ell}=F_{n\ell}(v)=\min\limits_{r>0}\left\{K_{n\ell}^{(f)}(r)+v f(r)\right\}
\end{equation}
For the power-law potentials $f(r)=r^q$, it is known by mean of simple scaling argument that the spectrum of the pure-power Hamiltonian satisfies
\begin{equation}\label{eq19}
-\Delta+v r^q\rightarrow F_{n\ell}^{(q)}(v)=E_{n\ell}^{(q)}(1)v^{2/(q+2)}.
\end{equation}
In order to compute the kinetic potentials $\overline f_{n\ell}(s)$, we notice from the minimization process of (\ref{eq14}) that $\overline f_{n\ell}^\prime(s)=-v^{-1}$ and consequently we have 
\begin{equation}\label{eq20}
s=F_{n\ell}^{(q)}(v)-v \overline f_{n\ell}(s)\Rightarrow \overline f_{n\ell}(s)={d\over d v} F_{n\ell}^{(q)}(v)
\end{equation}
which implies using (\ref{eq19}) that
\begin{equation}\label{eq21}
\overline f_{n\ell}(s)={2\over q+2}v^{-{q\over q+2}}E_{n\ell}^{(q)}(1).
\end{equation}
On the other hand, we have from the l.h.s. of (\ref{eq20}) that
\begin{equation}\label{eq22}
v^{-1}F_{n\ell}^{(q)}(v)=\overline f_{n\ell}(s)-s \overline f_{n\ell}^\prime(s)
\end{equation}
which implies using (\ref{eq21}) that
\begin{equation}\label{eq23}
\overline f_{n\ell}(s)={2\over q}\left({qE_{n\ell}^{(q)}\over q+2}\right)^{(q+2)/2}s^{-q/2}.
\end{equation}
The K functions are then computed by means of (\ref{eq16}) and (\ref{eq20})-(\ref{eq23}) to yields
\begin{equation}\label{eq24}
K_{n\ell}^{(f)}(r)=\left({2\over q}\right)^{2/q}\left({qE_{n\ell}^{(q)}\over q+2}\right)^{(q+2)/q}{1\over r^2}=\left({P_{n\ell}(q)\over r}\right)^2
\end{equation}
where we have defined 
\begin{equation}\label{eq25}
P_{n\ell}(q)=\left(E_{n\ell}^{(q)}\right)^{(2+q)/2q}\left[{2\over 2+q}\right]^{1/q}\left[{q\over 2+q}\right]^{1/2}
\end{equation}
The eigenvalues are then recovered by (\ref{eq18}) as
\begin{equation}\label{eq26}
E_{n\ell}=\min\limits_{r>0}\{\left({P_{n\ell}(q)\over r}\right)^2
+v r^q\}
\end{equation}
In ordered to obtained a definite bound, Hall \cite{hall83} used interesting geometric interpretation in terms of envelopes. If the potential shape $f(r)=g(h(r))$ is a smooth transformation $g$ of a soluble potential $h$, then the kinetic potentials associated with $f(r)$ are given by
\begin{equation}\label{eq27}
f(r)=g(h(r))\rightarrow \overline{f}_{n\ell}(s)\approx g(\overline{h}_{n\ell}(s))
\end{equation}
and the corresponding $K$ functions satisfies
\begin{equation}\label{eq28}
K_{n\ell}^{(f)}=(g\circ \overline{h}_{n\ell})^{-1}\circ (g\circ h)=\overline{h}_{n\ell}^{-1}\circ h = K_{n\ell}^{(h)}
\end{equation}
Therefore
\begin{equation}\label{eq29}
f=g(h)\rightarrow K^{(f)}\cong K^{(h)}
\end{equation}
and the eigenvalue approximations are given by
\begin{equation}\label{eq30}
E_{n\ell}\approx \min\limits_{r>0}\{K_{n\ell}^{(h)}(r)+v f(r)\}
\end{equation}
in which $g$ no longer appears.   This expression yields upper or lower bounds depending, respectively, whether $g$ is concave or convex \cite{hall92}-\cite{hall2002}. For $f(r)=g(r^q)=\sum_{i = 1}^{k}a_ir^{q_i}$: since $q_i < q_{i+1},$ clearly $g$ is convex if $q = q_1$ (lower bound) and concave if $q = q_k$ (upper bound).  We therefore have, by using (\ref{eq30}) with $h(r) = r^q$,
\begin{equation}\label{eq31}
E_{n\ell}=\min\limits_{r>0}\left\{\left({P_{n\ell}(q)\over r}\right)^2
+v \sum_{i = 1}^{k}a_ir^{q_i}\right\}.
\end{equation}
Or, equivalently, and by a change in the minimization variable,
\begin{equation}\label{eq32}
 E_{n\ell}=\min\limits_{r>0}\left\{{1\over{r^2}} +v \sum_{i = 1}^{k}a_i\left({P_{n\ell}(q)r}\right)^{q_i}\right\}.
\end{equation}
With $v = 1,$ this equation yields (\ref{eq9}) and we obtain a lower bound if $q = q_1$ and an upper bound if $q = q_k.$ This complete the proof of the theorem. \qed
\subsection{Proof of Theorem B} 
\noindent The first part of theorem B was introduced \cite{hall92} to improve the lower bounds for the ground state energy obtained in theorem A. The second part is  based on the Barnes et al's \cite{barnes} general lower-bound formula for the lowest eigenvalue of the Schr\"odinger operator $H=-\Delta+V(r)$ in $d\geq 1$ spatial dimensions. The extension to the potential sums, such as that of (\ref{eq1}), was introduced in \cite{hall65} where a detailed proof of theorem B can be found. \qed
\section{Fractional anharmonic oscillator} 
Before we study specific problems in quantum mechanics, we first consider the application of theorem A and Theorem B to the class of arbitrary fractionally anharmonic oscillator Hamiltonians \cite{znojil2}:
\begin{equation}\label{eq33}
H=-\Delta+\sum_{\delta\in Z}g_\delta r^{\delta}
\end{equation}
where $Z$ is an arbitrary finite set of the integer or rational numbers and the coupling $g_\delta,\delta\in Z$ are chosen so that the Hamiltonian supports the existence of a discrete  spectrum. It is known \cite{znojil2}-\cite{gerdt} that this class of Hamiltonian possesses elementary solutions for certain particular cases of the coupling $g_\delta$. For consistency, we assume $\delta\geq 2$, although the conclusion of theorems A and B are perfectly applicable for all $\delta\geq-1$, where, for example, the $P$ number in the case $\delta=-1$ is $P_{n\ell}^{d\geq 2}(-1)=(n+l+d/2-3/2)$. This class of Hamiltonian is a generalization of the Hamiltonian
\begin{equation}\label{eq34}
H=-\Delta+Cr^\alpha+Dr^\beta, \quad\beta>\alpha>0
\end{equation}
which has been used in the theory of heavy quarkonia \cite{francisco}-\cite{Quigg}. Denote $q=\min\limits_{\delta\in Z}\{\delta\}$ and $Q=\max\limits_{\delta\in Z}\{\delta\}$. By using theorem A, we immediately find analytic expressions for lower bounds $\epsilon_{n\ell}^{d}$ and upper bounds $E_{n\ell}^{d}$ for the eigenvalues of the Hamiltonian (\ref{eq33}): these can be written explicitly as 
\begin{equation}\label{eq35}
\epsilon_{n\ell}^{d}(\delta)\approx \min_{r > 0}\left[\frac{1}{r^2} + \sum\limits_{\delta\in Z \atop q=\min\limits_{\delta\in Z}\{\delta\}}g_\delta (P(q)r)^{\delta}
\right],
\end{equation}
and
\begin{equation}\label{eq36}
E_{n\ell}^{d}(\delta)\approx \min_{r > 0}\left[\frac{1}{r^2} + \sum\limits_{\delta\in Z \atop Q=\max\limits_{\delta\in Z}\{\delta\}}g_\delta (P(Q)r)^{\delta}
\right].
\end{equation}
Here the numbers $P(q)$ and $P(Q)$ are computed numerically by means of Eq.(\ref{eq3}) for rational $q,Q\neq -1,2$ by the use of direct numerical integration of the corresponding Schr\"odinger equations $(-\Delta +r^q)\psi=E_q\psi$ and $(-\Delta +r^Q)\psi=E_Q\psi$. An interesting improvement for the eigenvalues $\epsilon_{1\ell}^{d}(\delta)$ and $E_{1\ell}^{d}(\delta)$ can be obtained through the application of theorem B. The cost, however, is that the exact eigenvalues of the rational power-law potentials $V(r)=r^\delta$ for {\it each} $\delta\in Z$ must be computed numerically. Less accurate bounds can be obtain directly using the explicit $P$ numbers (\ref{eq11}) and (\ref{eq12}).  An important class \cite{znojil2} of the fractional anharmonic oscillator Hamiltonians (\ref{eq33}) that have found many applications \cite{znojil2} in quantum field theory \cite{Car} is given by 
\begin{equation}\label{eq37}
H=-\Delta+V(r)=-\Delta+\sum_{j=1}^{2q+1}g_j r^{2j},\quad g_{2q+1}=a^2>0.
\end{equation}
This class of Hamiltonians has found many applications not only in quantum mechanics (where $V(r)$ represents \cite{znojil2} an arbitrary potential in the limit $q\rightarrow \infty$)  but also, ofr example, in the Reggeon field theorem on the lattice \cite{fulco}. Theorem A gives immediate lower and upper bounds to the eigenvalues of (\ref{eq37}) as:
 \begin{equation}\label{eq38}
\epsilon_{n\ell}^{d}(\delta)\approx \min_{r > 0}\left[\frac{1}{r^2} + \sum\limits_{j=1}^{2q+1}g_j (P_{n\ell}^{d}(2)r)^{2j}
\right].
\end{equation}
and
\begin{equation}\label{eq39}
E_{n\ell}^{d}(\delta)\approx \min_{r > 0}\left[\frac{1}{r^2} + \sum\limits_{j=1}^{2q+1}g_j (P_{n\ell}^{d}(2q+1)r)^{2j}
\right].
\end{equation}
where $P_{n\ell}^{d}(2)$ is given by Eq.(\ref{eq8}) and $P_{n\ell}^{d}(2q+1)$ is given by (\ref{eq3}), respectively. 

\section{Quantum anharmonic oscillator} 
\noindent In this section, we consider the Schr\"odinger equation
\begin{equation}\label{eq40}
\left(-\omega\Delta+a r^2+b r^{2m}\right)\psi=E(\omega,a,b)\psi,\quad\quad m=2,3,4,\dots
\end{equation}
where $\omega,~a$ and $b$ are positive parameters and the potential in (\ref{eq33}) is a single-well potential which describes for $m=2,3,4,5,\dots$ the quartic, sextic, octic, and decadic oscillators, and so on. It is easy to check that for the energy in (\ref{eq40}) the following scaling relation holds
\begin{eqnarray}\label{eq41}
E(\omega,a,b)&=&(a \omega)^{1/2}{E}\left(1,1,{b\omega^{(m-1)/2}\over 
a^{(m+1)/2}}\right),\nonumber\\ 
\psi(r; \omega,a,b)&=&\psi\left(\left({a\over\omega}\right)^{\frac14} r; 1,1,{b\omega^{(m-1)/2}\over 
a^{(m+1)/2}}\right) .
\end{eqnarray}
Thus the original problem (\ref{eq40}) is essentially a single-parameter problem which we now write as
\begin{equation}\label{eq42}
H^{(m)}\psi=\left(-\Delta +r^2+\lambda r^{2m}\right)\psi=E(\lambda)\psi ,\quad\quad m=2,3,4,\dots
\end{equation}
where $E(\lambda)={E}(1,1,\lambda)$ and $\lambda={b\omega^{(m-1)/2}\over 
a^{(m+1)/2}}$.  The Schr\"odinger equation with the quantum anharmonic oscillators (\ref{eq42}) are among the most widely studied models in quantum mechanics. In spite of their simplicity, they give rise to interesting problems, both computationally and conceptually \cite{weniger}. A rigorous analysis of the mathematical properties of the anharmonic oscillator Hamiltonians $H^{(2)}$ was made by Simon \cite{simon} and by the seminal work of Bender and Wu \cite{bender}. The aim in the discussion we present below is to derive simple upper- and lower-bound formulae based on Theorems A and B. For the anharmonic oscillator potentials
\begin{equation}\label{eq43}
f(r)=r^2+\lambda r^{2m},\quad m=2,3,\dots
\end{equation}
Theorem A implies that  
\begin{equation}\label{eq44}
{\cal E}(\lambda)\approx \min_{r > 0}\left[\frac{1}{r^2} + \alpha r^2+\lambda \beta r^{2m}\right].
\end{equation}
where
\begin{itemize} 
\item ${\cal E}(\lambda) \leq E(\lambda)$ is an lower bound, if $\alpha=(P_{n\ell}(2))^2$ and $\beta=(P_{n\ell}(2))^{2m}$.
\item ${\cal E}(\lambda) \geq E(\lambda)$ is an upper bound, if $\alpha=(P_{n\ell}(2m))^2$ and $\beta=(P_{n\ell}(2m))^{2m}$.
\end{itemize}
\noindent Furthermore, Theorem B implies that if $\alpha=(P_{n\ell}(2))^2$ and $\beta=(P_{n\ell}(2m))^{2m}$, then ${\cal E}(\lambda) \leq E(\lambda)$ for $n=1$, and ${\cal E}(\lambda) \approx E(\lambda)$ for all $n\geq 2$.
Let $x=r^2$, we note that the minimization of (\ref{eq44}) occurs at $-{1\over x^2}+\alpha+m\lambda \beta x^{m-1}=0.$ Multiplying through by $x$ and solving for $\lambda \beta x^{m}$, we can easily  show that the minimization of (\ref{eq44}) occurs at
\begin{equation}\label{eq45}
r^2={mE -\sqrt{E^2m^2-4\alpha(m^2-1)}\over 2\alpha(m-1)}
\end{equation}
and consequently we have
\begin{equation}\label{eq46}
\lambda={1\over \beta}\left({2\alpha(m-1)\over mE-\delta}\right)^m\left({4\alpha\over (mE-\delta)}-{E\over (m-1)}\right)
\end{equation}
where
$\delta={\sqrt{E^2m^2-4\alpha(m^2-1)}}$. Thus for finding the energy eigenvalues of anharmonic-oscillator Hamiltonians $H^{(m)}$ in (\ref{eq42}) one has to solve Eq.(\ref{eq46}) for the given $\lambda$. It is clear that at $\lambda=0$, equation (\ref{eq46}) implies $E=2\sqrt{\alpha}$, with $\alpha=(P_{n\ell}^{(q)}(2))^2=(2n+l+{d\over 2}-2)^2$, as given by (\ref{eq8}). Consequently, $E=4n+2l+d-4$, the result for the $d$-dimensional harmonic oscillator \cite{movr}. 
When equation (\ref{eq46}) is used to determine the lower or the upper bounds to the exact eigenvalues of the $\lambda r^{2m}$ oscillator, it is clear that the formula is expressed in terms of a single constant for any value of $\lambda$. This follows from the fact that, for lower or upper bound, $\alpha^m=\beta$ and equation (\ref{eq46}) reduces to
\begin{equation}\label{eq47}
\lambda={2^m(m-1)^{(m-1)}\over (m+1)}{(-E+\sqrt{m^2E^2-4\alpha(m^2-1)})\over (mE-\sqrt{m^2E^2-4\alpha(m^2-1)})^m}.
\end{equation}
This is a remarkable simple formula that gives a global lower and upper bound to the exact eigenvalues for a given $\lambda$ for all $n=1,2,\dots$ and $l=0,1,2,\dots$ in $d$ dimensions,  accordingly as $\alpha=(2n+l+{d\over 2}-2)^2$ and $\alpha=(P_{n\ell}^{(d)}(2m))^{2}$, respectively. In particular, a global formula that gives a lower bound for all $n=1,2,\dots$, $m=2,3,\dots$ is 
\begin{eqnarray}
\lambda&=&{2^m(m-1)^{(m-1)}\over (m+1)}{(-E+\sqrt{m^2E^2-4(2n+l+{d\over 2}-2)^2(m^2-1)})\over (mE-\sqrt{m^2E^2-4(2n+l+{d\over 2}-2)^2(m^2-1)})^m}.\nonumber\\ \label{eq48}
\end{eqnarray}
Note in the case of $d=1$, we should set either $l=-1$ or $l=0$ to obtain a lower bound to  even or odd (exact) eigenvalues, respectively. Despite the generality of (\ref{eq47}), we should like to make two immediate remarks concerning the application of theorem A: (i) Formula (\ref{eq47}), in general, gives a loose bound; (ii) the upper bound $\alpha=(P_{n\ell}^{(d)}(2m))^{2},m=2,3,\dots$ which is obtained by means of Eq.(\ref{eq4}), requires knowledge of the exact eigenvalues of Schr\"odinger equation $(-\Delta_d+r^{2m})\psi=\epsilon_{n\ell}^{(d)}(2m)\psi$. In this paper we have found the values of $P_{n\ell}^{(d)}(2m)$ by the numerical integration of the Schr\"odinger equation just mentioned, and then we used Eq.(\ref{eq4}) to find the corresponding $P$-numbers. For immediate use of equations (\ref{eq47}) and (\ref{eq48}), we report in Table 1 the values of $P_{10}^{(1)}(2m)$ and $\beta=(P_{n\ell}^{(1)}(2m))^{2m}$ for different values of $m$.
\begin{center}
{{Table~1}: Values of $P_{10}^{(1)}(2m)$ and $\beta=(P_{n\ell}^{(1)}(2m))^{2m}$ for different $m$.}
\vskip0.1in
\noindent\begin{tabular}{|c|c|c|}
\hline
$m$& $P_{10}^{(1)}(2m)$ &  $\beta$\\
\hline
2&~0.648~283~101~647~721~4~&~0.176~627~696~530~967~9~\\
\hline
3&~0.752~213~287~729~753~3~&~0.181~153~198~043~223~7~\\
\hline
4&~0.830~692~879~447~472~3~&~0.226~737~786~349~046~1~\\
\hline
5&~0.892~746~975~167~740~8~&~0.321~576~181~371~282~8\\
\hline
6&~0.943~407~187~840~825~1~&~0.497~038~660~113~318~0\\
\hline
\end{tabular}
\end{center}
In order to illustrate the above discussion, we consider the case of finding the eigenvalues of the anharmonic oscillator Hamiltonian $-{d^2\over dr^2}+r^2+0.01 r^4$. Eq.(\ref{eq48}) gives a lower bound 1.00248 and Eq.(\ref{eq47}) with $\alpha=(P_{10}^{(1)}(2m))^2$ gives an upper bound of 1.32038. The exact eigenvalue in this case reads 1.00737. In order to improve these bounds, we can make use of Theorem B.  For the ground state eigenvalues in $d\geq 1$, the first part of Theorem B can be applied to obtain a a more accurate lower bound formula. Further, the second part of theorem B   can be used to obtain straightforward lower and upper bound without the necessity of the numerical computation of $P$-numbers, thanks to the explicit approximate values of $P_{10}^{(d)}(2m)$ given by (\ref{eq11}) and (\ref{eq12}). In either case, equation (\ref{eq46}) gives a simple general formula for the energy bound of $E=E(\lambda)$ with reasonable accuracy 
\begin{equation}\label{eq49}
{1\over 2^m} {(m-1)^{(m-1)}\over (m+1)} {(-E+\sqrt{m^2(E^2-d^2)+d^2})\over 
(mE-\sqrt{m^2(E^2-d^2)+d^2})^m}=\beta \lambda,
\end{equation}
for $m=2,3,\dots$ where now if $\beta=(P_{10}^{(d)}(2m))^{2m}$ as given by Table (1), Eq.(\ref{eq48}) gives a lower bound for given $\lambda$. The results of this formula is illustrated in the last columns of Tables 2 and 3. On other hand, if $\beta$ is a fixed number given by (\ref{eq11}) and (\ref{eq12}), then Eq.(\ref{eq49}) gives lower and upper bound, respectively.  Note that Theorem B, still allows us to conclude that Eq.(\ref{eq49}) yields a reasonable approximation to the excited-state energies for $n\geq 2$.  However, in this case $\beta=(P_{n0}^{(d)}(2m))^{2m}$ is strictly given by (\ref{eq4}). In the case of $d=1$, Eq.(\ref{eq48}) reads ($m=2,3,\dots$)
\begin{equation}\label{eq50}
{1\over 2^m} {(m-1)^{(m-1)}\over (m+1)} {(-E+\sqrt{m^2(E^2-1)+1})\over 
(mE-\sqrt{m^2(E^2-1)+1})^m}=\beta \lambda,
\end{equation}
where  $\beta=(P_{10}^{(1)}(2m))^{2m}$ is given by (\ref{eq11}) and (\ref{eq12}) for a lower and an upper bound respectively. Equation (\ref{eq50}) can be compared with the recent formula introduced by Bhattacharya et al \cite{ran} for the approximate ground state energy of  the Hamiltonian $-\Delta + r^2 +\lambda r^{2m}$ in one dimension, namely 
\begin{equation}\label{eq51}
(E^{(m)})^{(m+1)}-(E^{(m)})^{(m-1)(1+2/(m+2+\lambda))}=(K_0^{(m)})^{(m+1)}\lambda
\end{equation}
where $K_0^{(2)}=1.06036209$, $K_0^{(3)}=1.14480245$, $K_0^{(4)}=1.22582011$ are the coefficients of the first terms in the respective strong coupling expansion computed by Weniger \cite{weniger}. In Tables 2 and 3, we have compared our lower and upper bounds given by (\ref{eq49}), for the quartic and sextic anharmonic oscillator, along with the exact eigenvalues obtained by the direct integration of the corresponding Schr\"odinger equation.
We have also compared the best lower bound $E_L$ obtained using (\ref{eq49}), where $\beta$-values were given by table 1, with the approximate eigenvalues of the ground state energy $E_b$ computed by Bhattacharya et al \cite{ran} using formula (\ref{eq50}). 
Recently, Dasgupta et al \cite{dasgupta} have extend the work of Bhattacharya et al \cite{ran} to evaluate the excited state energies, still in the one-dimensional case. As in the case of the ground state (\ref{eq50}), they found that the excited-state energies for the $\lambda r^{2m}$ oscillator defined by the one-dimensional Hamiltonian operator $H=-d^2/dr^2+r^2+\lambda r^{2m}$ are also a polynomial equation of the same degree and are given by
\begin{equation}\label{eq51}
\bigg({E^{(m,n)}\over 2n+1}\bigg)^{(m+1)}-\bigg({E^{(m,n)}\over 2n+1}\bigg)^{(m-1)}=(K_0^{(m,n)})^{(m+1)}\lambda
\end{equation}
where $E^{(m,n)}$ is the $n^{th}$ excited state energy of the $\lambda r^{2m}$ oscillator and $K_0^{(m,n)}$ are constants \cite{dasgupta}. Our formulas (\ref{eq46}) and (\ref{eq47}) are more general and seems to yield more accurate results, even for large values of the coupling parameter $\lambda$.
\section{Conclusion}
The application of envelope theory and kinetic-potential techniques to polynomial potentials has yielded fairly general and good energy bounds for arbitrary values of the coupling constants. As specific examples, the application of theorems A and B to the quantum anharmonic oscillators has produced a global energy formula sufficient to generate all energy levels in arbitrary dimension for $r^{2m}$ anharmonic oscillators with a fair degree of accuracy.  The main emphasis of this paper has been on energy formulas that are also bounds. However the energy formula of Theorem B (ii), namely
$$
E \approx \min_{r > 0}\left[\frac{1}{r^2} + \sum_{i = 1}^{k}a_i\left(P_{n\ell}^{(d)}(q_i)r\right)^{q_i}\right],
$$
which does indeed yield a lower bound for the bottom of spectrum $(n = 1)$ in each angular-momentum subspace, is a remarkably general and accurate approximation: it requires the input of the pure-power P numbers, and it then predicts approximately, for all the eigenvalues in all dimensions, how the spectrum generated by the potential sum depends on the mixing parameters $\{a_i\};$ it also has the attractive collocation property that it is exact whenever all but one of the potential coefficients are zero.
\section*{Acknowledgments}
\medskip
\noindent Partial financial support of this work under Grant Nos. GP3438 and GP249507 from the 
Natural Sciences and Engineering Research Council of Canada is gratefully 
acknowledged by two of us ([RLH] and [NS]).
\newpage

\noindent{Table~2}: {\it Calculated values of upper and lower bounds, using (\ref{eq43}), to the ground state energy of the quartic anharmonic oscillator, along with exact values, for different values of $\lambda$. The comparison between the lower bound $E_L$ given by Eq.(\ref{eq43}) using the exact values of $\beta$ by means of table (1), and the approximate eigenvalues $E_b$ of Bhattacharya et al \cite{ran} using (\ref{eq44}) are also shown.}
\vskip0.1in
\begin{center}
\noindent\begin{tabular}{|c|c|c|c|c|c|}
\hline
$\lambda$& Exact value &  Lower bound  &  Upper bound  & $E_b$ & $E_{L}$ \\
~&~&using Eq.(\ref{eq43})&using Eq.(\ref{eq43})& Ref. \cite{ran} & using Eq.(\ref{eq43})\\
\hline
0.001&1.00075& 1.00062& 1.00075& 1.00079 & 1.00071\\
\hline
0.01&1.00737& 1.00614& 1.00739& 1.00783 & 1.00697\\
\hline
0.1&1.06529& 1.05585& 1.06620& 1.07005& 1.06275\\
\hline
0.2&1.11829& 1.10288& 1.12062& 1.12702& 1.11473\\
\hline
1.0&1.39235& 1.35510& 1.40332& 1.41155& 1.38754\\
\hline
4.0&1.90314& 1.83699& 1.92881& 1.91489& 1.89895\\
\hline
10.0&2.44917& 2.35648& 2.48862& 2.45005 & 2.44575\\
\hline
50.0&4.00399& 3.841639& 4.078522& 3.99621& 4.00182\\
\hline
100.0&4.99942& 4.79395& 5.09516& 4.99161&4.99766\\
\hline
1000.0&10.63979& 10.19449& 10.85151& 10.63521&10.63896\\
\hline
2000.0&13.38844& 12.82706& 13.65591& 13.38474& 13.38778\\
\hline
\end{tabular}
\end{center}
\vskip0.3in
\noindent{Table~3}: {\it Calculated values of upper and lower bound to the ground state energies of the sextic anharmonic oscillator along with exact values for different values of $\lambda$. The comparison between the lower bound $E_L$ given by Eq.(\ref{eq43}) and the approximate eigenvalues $E_b$ of Bhattacharya et al \cite{ran} using (\ref{eq44}) are also shown.}
\vskip0.1in
\begin{center}
\noindent\begin{tabular}{|c|c|c|c|c|c|}
\hline
$\lambda$& Exact value &  Lower bound  &  Upper bound  & $E_b$ &$E_L$\\
~&~&using Eq.(\ref{eq43})&using Eq.(\ref{eq43})& Ref. \cite{ran} & using Eq.(\ref{eq43})\\
\hline
0.001&1.00185& 1.000932& 1.001859& 1.00143& 1.00144\\
\hline
0.01&1.01674& 1.008994& 1.017387& 1.01374 & 1.01366\\
\hline
0.1&1.10908& 1.070681& 1.119935& 1.10565& 1.09920\\
\hline
0.2&1.17389& 1.119782& 1.192805& 1.17513& 1.16261\\
\hline
1.0&1.43653& 1.334560& 1.484050& 1.44870& 1.42400\\
\hline
4.0&1.83044& 1.675050& 1.916177& 1.83193 & 1.82058\\
\hline
10.0&2.20572& 2.004582& 2.322916& 2.19235& 2.19734\\
\hline
50.0&3.15902& 2.850163& 3.348809& 3.13471&3.15304\\
\hline
100.0&3.71698& 3.347427& 3.946987& 3.69348& 3.71187\\
\hline
1000.0&6.49235& 5.828630& 6.914382& 6.47694&6.48941\\
\hline
2000.0&7.70174& 6.911387& 8.205757& 7.68861&7.69925\\
\hline
\end{tabular}
\end{center}

\end{document}